\documentclass{article}
\usepackage{fleqn}
\usepackage{epsfig}
\usepackage{amsmath,amssymb}

\begin{document}

\begin{center}
{\bf\Large The complete relativistic kinetic model of violation of
symmetry in isotopic expanding plasma and production of baryons in
hot Universe. II. Numerical model: X-boson distribution function.}\\[12pt]
Yu.G.Ignat'ev, K.Alsmadi\\
Kazan State Pedagogical University,\\ Mezhlauk str., 1, Kazan
420021, Russia
\end{center}

\begin{abstract}
In terms of proposed by authors general-relativistic kinetic model
of baryon production in expanding primordially symmetrical hot
Universe calculates distribution function of extra-massive bosons
and concerned with it variables.\end{abstract}

\section{Introduction}
In previous paper \cite{Yu_barI} in terms of theory , developed
earlier by one of the authors,\footnote{see Yu.G.Ignat'ev,
\cite{Ignat4}, \cite{Ignat5}} was proposed an accurate kinetic
model of baryon production in primordially baryon-symmetrical
Universe on basis of field model of interaction of particles of
type SU(5). Here and further references to formulas of authors
previous paper \cite{Yu_barI} are indicate as (I,N), where N -
number of formula of stated paper.

\section{Weak violence of charge symmetry in standard SU(5) model}

Using weakness of violence of $CP$-symmetry
$$\Delta r=r-\bar{r}\lesssim 10^{-6}$$
and thereby smallness of chemical potentials ($\lambda_a \lesssim
\Delta r \ll 1$), we can reduce referred above system of
integro-differential equations to a system of linear
integro-differential equations. In present paper we will consider
simple model of baryogenesis, in which $CP$-invariance is violated
in decay channels of only one type of bosons, $X$-boson. In this
case equations (I.40), (I.42) assume following form:%
\begin{eqnarray}
\dot{f}_X+f_X\frac{a(t)S^2}{2\pi\sqrt{a^2(t)m^2_X+\mathbb{P}^2}}\times\hskip 3cm\nonumber\\
\times\left[\bar{r}\beta(d)+ (1-\bar{r})\beta(-u)\right]=\nonumber\\
\frac{a(t)S^2}{2\pi\sqrt{a^2(t)m^2_X+\mathbb{P}^2}}\times\hskip 3cm\nonumber\\
\left[\bar{r}\beta(d)f^0_X(2d)+(1-\bar{r})\beta(-u)f^0_X(-2u)\right];\label{III.116}
\end{eqnarray}

\begin{eqnarray}
\Delta \dot{\mathcal{N}}_\alpha=\frac{2a(t)S^2N_X}{3(2\pi)^3}
\int\limits_0^\infty\frac{\mathbb{P}^2d\mathbb{P}}
{\sqrt{a^2(t)m^2_X+\mathbb{P}^2}}\times\nonumber\\
\left\{(1-\bar{r})\beta(u)[f_{\bar{X}}-f^0_X(2u)]-\right.\nonumber\\
\left.-(1-r)\beta(-u)[f_X-f^0_X(-2u)]\right\};\label{III.117}
\end{eqnarray}

\begin{eqnarray}
\Delta \dot{\mathcal{N}}_\kappa=\frac{2a(t)S^2N_X}{3(2\pi)^3}
\int\limits_0^\infty\frac{\mathbb{P}^2d\mathbb{P}}
{\sqrt{a^2(t)m^2_X+\mathbb{P}^2}}\times\nonumber\\
\left\{r\beta(d)[f_{\bar{X}}-f^0_X(2d)]-
\bar{r}\beta(-d)[f_{\bar{X}}-f^0_X(-2d)]\right\}.\nonumber\\\label{III.118}
\end{eqnarray}
Here separately wrote out equations for anoquarks-
(\ref{III.117}), and katoquarks -(\ref{III.118}). At derivation of
these equation was allowed that:
\begin{equation}\label{III.118a}
\beta(x,x)=2\beta(x).
\end{equation}

Let us write out formulas for coefficients that belong to these
equations in linear in $\Delta r, \lambda$ approximation. In zero
approximation:
\begin{equation}\label{III.118b}
\beta(0)=\beta_0(\mathbb{P},t),
\end{equation}
where:
\begin{equation}\label{III.118c}
\beta_0(\mathbb{P},t)= \displaystyle{\frac{T}{p} \ln
\frac{1+\exp(p_+/T)}{\exp(p/2T)[1+\exp(p_-/T)]}}
\end{equation}
and it is necessary to substitute:
\begin{equation}\label{III.118d}
\frac{p_{\pm}}{T}=\frac{1}{2}\frac{\sqrt{a^2(t)m^2_X+\mathbb{P}^2}\pm
\mathbb{P}}{\mathcal{T}}.
\end{equation}

Further:
\begin{equation}\label{III.118e}
f^0_X(\xi)=f^0_X(0;\mathbb{P},t)+\xi\;\frac{\exp(E_X/T)}{\left[\exp(E_X/T)-1\right]^2},
\end{equation}
where:
\begin{equation}\label{III.118g}
f^0_X(0;\mathbb{P},t)=\left[\exp(E_X/T)-1\right]^{-1}
\end{equation}
and
$$\frac{E_X}{T}=\frac{\sqrt{a^2(t)m_X^2+\mathbb{P}^2}}{\mathcal{T}}.$$

Integrating kinetic equations for  X-bosons (\ref{III.116}) with
initial conditions (I.43) in linear in $\Delta r, \lambda$
approximation we will determine:
\begin{equation}\label{III.119}
f_X(\mathbb{P},t)=f_X^0(0;\mathbb{P},0)+\delta
f_0(\mathbb{P},t)+\delta f(\lambda),
\end{equation}
where $\delta f(\lambda)$ - linear in $\lambda$ functional, hence:
\begin{equation}\label{III.120}
\delta(-\lambda)=-\delta(\lambda),
\end{equation}
and it doesn't contribute in final output; and $\delta f$ -
deviation from equilibrium of distribution function:
\begin{equation}\label{III.121}
\delta f(\mathbb{P},t)=- e^{-\Phi(\mathbb{P},t)}\int\limits_0^t
e^{\Phi(\mathbb{P},t')}\dot{f}_0(0;\mathbb{P},t')dt'
\end{equation}
- deviation from equilibrium of boson distribution function in
symmetrical plasma $\lambda=0$, and incorporated notation:
\begin{equation}\label{III.122}
\Phi(\mathbb{P},t)=\frac{1}{\tau_0}\int\limits_0^{t}
\frac{a(t')\beta_0(\mathbb{P},t')d
t'}{\sqrt{a^2(t')+\mathbb{P}^2/m_X^2}};
\end{equation}
$\tau_0$ - proper time of free X-boson's decay:
\begin{equation} \label{III.123}
\tau_0=\frac{4\pi m_X}{s^2} \sim \frac{3}{2}(m_X \alpha)^{-1}.
\end{equation}

Since fermions lay in condition of ËÒÐ and are ultrarelativistic,
their concentrations are equal:
\begin{equation}\label{III.119a}
\mathcal{N}=\frac{1}{\pi^2}\int\limits_0^\infty
\frac{1}{\exp(-\lambda+\mathbb{P}/\mathcal{T})-1}\mathbb{P}^2d\mathbb{P}.
\end{equation}
In conditions of weak violence of $CP$ - invariance, when
$\lambda_a\ll 1$ we will receive from here approximately,
separating at degrees of smallness $\lambda$ :
$$\mathcal{N}\approx\frac{3}{2} \frac{\mathcal{T}^3 \zeta(3)}{\pi^2}+
\lambda \frac{\mathcal{T}^3}{6} \Rightarrow$$
\begin{equation}\label{III.119b}
\Delta N_a \approx \lambda_a \frac{\mathcal{T}^3}{3}.
\end{equation}

In standard SU(5) model probabilities of X-boson's decay in
lepton, ($qe$), and quark, ($\bar{q}\bar{q}$), channels are the
same \footnote{subject to colors and charms}, i.e.:
\begin{equation}\label{III.123a}
1-2r =O^1(\Delta r).
\end{equation}
Subject to this factor and relation (\ref{III.119b}) {\it in
standard} SU(5) model from equations for anoquarks-
(\ref{III.117}), and katoquarks -(\ref{III.118}) we can receive
one closed first-order equation on variable $B=u+2d$, i.e., on
excessive concentration of baryons:
$$\Delta \mathcal{N}_B=\frac{1}{3}\mathcal{N}_B \mathcal{T}^3:$$

$$\frac{d}{dt}\Delta \mathcal{N}_B+\Delta \mathcal{N}_B\frac{2
N_X}{\pi^2 \mathcal{T}^3}\int\limits_0^\infty
\mathbb{P}^2\dot{\Phi}f_0 \beta_0 d\mathbb{P}=$$
\begin{equation}\label{III.125}
=\frac{2\Delta r N_X}{3\pi^2}\int\limits_0^\infty
\mathbb{P}^2\dot{\Phi}\beta_0 \delta f_0 d \mathbb{P},
\end{equation}
integrating which subject to initial conditions (I.43), we will
obtain:
\begin{equation}\label{III.126}
\Delta \mathcal{N}_B(\infty)=\frac{4 \Delta r
N_X}{3\pi^2}\int\limits_0^\infty \exp\left(-\int\limits_t^\infty
\Psi(t')dt'\right)G(t)dt,
\end{equation}

where
\begin{equation}\label{III.127}
\Psi(t)=\frac{2N_X}{\pi^2 \mathcal{T}^3}\int\limits_0^\infty
\mathbb{P}^2\dot{\Phi}f_0 \beta_0 d\mathbb{P},
\end{equation}
\begin{equation}\label{III.128}
G(t)=\frac{1}{\pi^2}\int\limits_0^\infty \mathbb{P}^2
\dot{\Phi}\delta f d \mathbb{P}.
\end{equation}
Now it remains to calculate variable $\delta_S=\delta n_B/S$,
where entropy density of ultrarelativistic gas is equal:
\begin{equation}\label{III.128}
S=\frac{2\pi^2}{45}NT^3\Rightarrow
\mathcal{S}=\frac{2\pi^2}{45}N\mathcal{T}^3.
\end{equation}

By that the task is formally solved. We should note that in
contrast to papers \cite{Fry1}-\cite{Fry3}, in which distribution
function of $X$-bosons was modelled by quasi-hydrodynamic
distribution and was obtained by method of numerical integration
of kinetic equations, distribution function of X-bosons in present
article obtains by strict integration of kinetic equations and
$B(\infty)$ determines in quadratures:
\begin{equation}\label{128a}
\delta_S=\frac{\Delta \mathcal{N}_B}{\mathcal{S}}.
\end{equation}

\section{Analysis of solution}

Further we will assume that firstly X-bosons decay in general on
intermediate stages of expansion, when  $T\sim m_X$, and, secondly
that part of X-bosons sufficiently small in comparison with
general number of particles: $N_X\ll N$, so with enough accuracy
grade we can assume cosmological plasma on decay stage of X-bosons
is ultrarelativistic. This gives us laws of variation of
dimensioned factor and temperature in due course:
\begin{equation}\label{yunew0}
a(t)=a_0\sqrt{t}; \quad T=\mathcal{T}_0\frac{1}{\sqrt{t}},
\end{equation}
where $\mathcal{T}_0$ (ñì. (I.44)) is plasma's temperature in
Planck units on Planck moment of time:
\begin{equation}\label{yunew0a}
\mathcal{T}_0=\displaystyle{\left(\frac{45}{16\pi^3
N}\right)^{1/4}}. \end{equation}

Let us explore obtained in previous section linear solution for
that we will come over from variables $t,p$ to new variables
$\eta,\xi$:
\begin{equation}\label{yunew1}
t=\tau_0 \eta;
\end{equation}

\begin{equation}\label{yunew2}
p=\frac{1}{\sqrt{\eta}}m_X\xi; \quad\Rightarrow
\mathbb{P}=m_X\xi\sqrt{\tau_0} .
\end{equation}
where $\tau_0$ - proper time of free X-boson's decay
(\ref{III.123}), in that way to valuation of dimensionless time
$\eta=1$ it corresponds time $t=\tau_0$. Let us incorporate
dimensionless parameter $\sigma$ \cite{Ignat5}, depending from
constants of field theory \footnote{Incorporated in \cite{Ignat5}
parameter $\sigma$ is equal to quadrate of $\sigma$, which uses in
present article.}:
\begin{equation}\label{III.129}
\sigma =
\frac{m_X}{T(\tau_0)}=\frac{m_X\sqrt{\tau_0}}{\mathcal{T}_0} =
\frac{\chi \sqrt{m_X}}{\sqrt{\alpha} \mathcal{T}_0},
\end{equation}
which is equal to relation of X-boson's mass to temperature at a
point of it's half-decay, where $\chi$ - dimensionless parameter
of 1 order, depending from parameters of field theory, $\alpha$ -
coupling constant (îò 0.1 äî 0.01).

Then we will receive expressions:
\begin{equation}\label{yunew3}
\frac{p}{T}=\sigma\xi; \quad \frac{E}{T}=\sigma\sqrt{
\eta+\xi^2};\quad \frac{p_\pm}{T}=\sigma(\sqrt{\eta+\xi^2}\pm \xi)
\end{equation}
\begin{equation}\label{yunew3a}
f_0(t,\mathbb{P})=\left[1+\exp(\sigma\sqrt{
\eta+\xi^2})\right]^{-1}
\end{equation}
$$\beta_0(\mathbb{P},t)=\beta_0(\xi,\eta,\sigma)= $$
\begin{equation}\label{yunew4}
=\displaystyle{\frac{1}{\xi\sigma} \ln
\frac{1+\exp(\sigma/2(\sqrt{\eta+\xi^2}+\xi))}
{\left[1+\exp(\sigma/2(\sqrt{\eta+\xi^2}-\xi))\right]\exp(\sigma\xi/2)}}
\end{equation}
 On fig. \ref{beta1} is shown relation of statistical factor $\beta_0(\xi,\eta,\sigma)$ (\ref{yunew4})
from parameters $\sigma$ è $\xi=p$.

\vskip 24pt\noindent \refstepcounter{figure}%\setcounter{figure}{1}
\epsfig{file=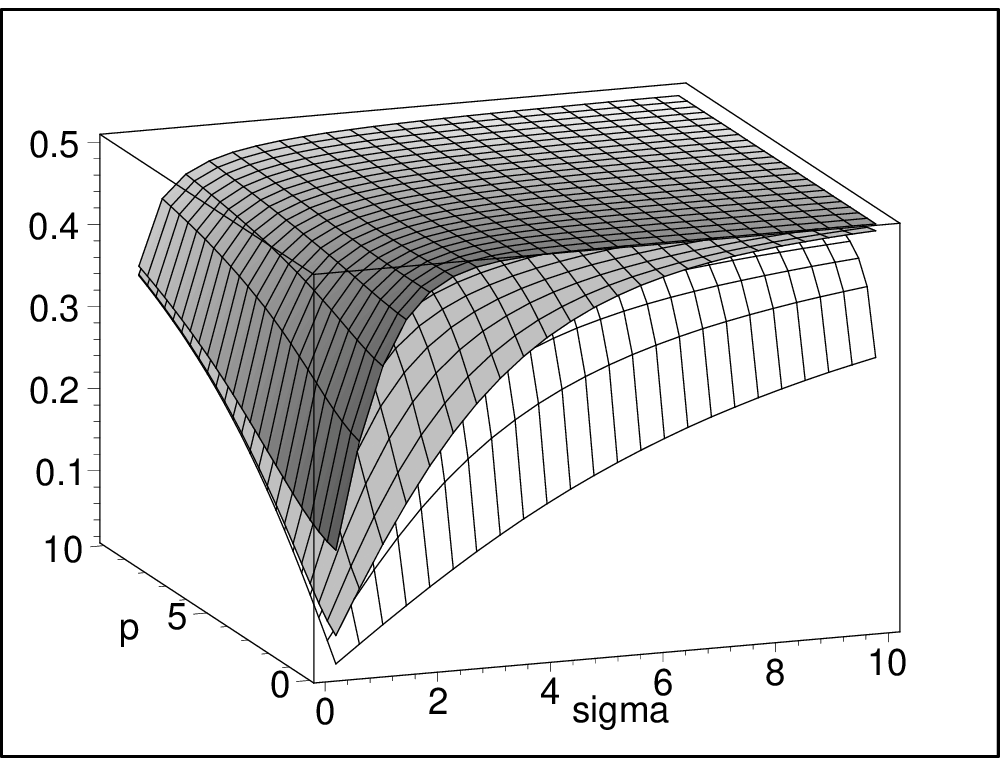,height=7cm,width=7cm}\label{beta1} \vskip
12pt \noindent {Fig.\bf \thefigure.}\hskip 12pt{\sl Function
$\beta_0(\xi,\eta,\sigma)$ for values of time $\eta=0,1,10$
(bottom-up ) .\hfill}
\vskip 12pt\noindent%

\subsection{Boltzmann distribution}
Let us calculate function $\Phi(\mathbb{P},t)$ (\ref{III.122}),
going over variables $\xi,\eta$ according to formulas
(\ref{yunew1}), (\ref{yunew2}):
\begin{equation}\label{yunew5_0}
\Phi(\mathbb{P},t)=\Phi(\xi,\eta)=\int\limits_0^\eta
\frac{\beta_0(\xi,\eta,\sigma))\sqrt{\eta'}d\eta'}{\sqrt{\eta+\xi^2}}.
\end{equation}
In Boltzmann approximation, substituting statistical factor
$\beta_0$ íà $\frac{1}{2}$ and executing integration we will
receive:
$$\Phi(\xi,\eta)=\frac{1}{8}(\sqrt{\eta}+\sqrt{\eta+\xi^2})^2-$$
\begin{equation}\label{yunew5_1}
-\frac{1}{8}\frac{\xi^4}{(\sqrt{\eta}+\sqrt{\eta+\xi^2})^2}
-\xi^2\ln\frac{(\sqrt{\eta}+\sqrt{\eta+\xi^2})^2}{\xi}.
\end{equation}

\subsection{Deviation of X-bosons from equilibrium}
Substituting function $\Phi(\xi,\eta)$ in form (\ref{yunew5_1})in
expression (\ref{III.121}) for X-boson's distribution function
deviation from equilibrium and going over variables
(\ref{yunew1}), (\ref{yunew2}), we will obtain:
\begin{equation}\label{III.121_new}\delta f(\xi,\eta,\sigma)=\frac{1}{2}\sigma
\exp(-\Phi(\xi,\eta))\int\limits_0^\eta
\frac{\exp\left(\Phi(\xi,\eta')+\sigma \sqrt{\eta'+\xi^2}\right)d\eta'}{%
[1+\exp(\sigma \sqrt{\eta'+\xi^2})]^2 \sqrt{\eta'+\xi^2}}.
\end{equation}
On Fig. \ref{delta-3}, \ref{delta-4} results of numerical
integration of expression (\ref{III.121_new}) for relative
deviation of distribution function from equilibrium, $\delta
f/f0$, subject to time, $\eta$, and momentum, $\xi$ for value
$\sigma=1$ are shown. Let us notice that fermi-distribution is
substituted to boltzmann distribution only in expression
(\ref{yunew5_0}) for function $\Phi(\xi,\eta)$).

\vskip 24pt\noindent \refstepcounter{figure}%\setcounter{figure}{1}
\epsfig{file=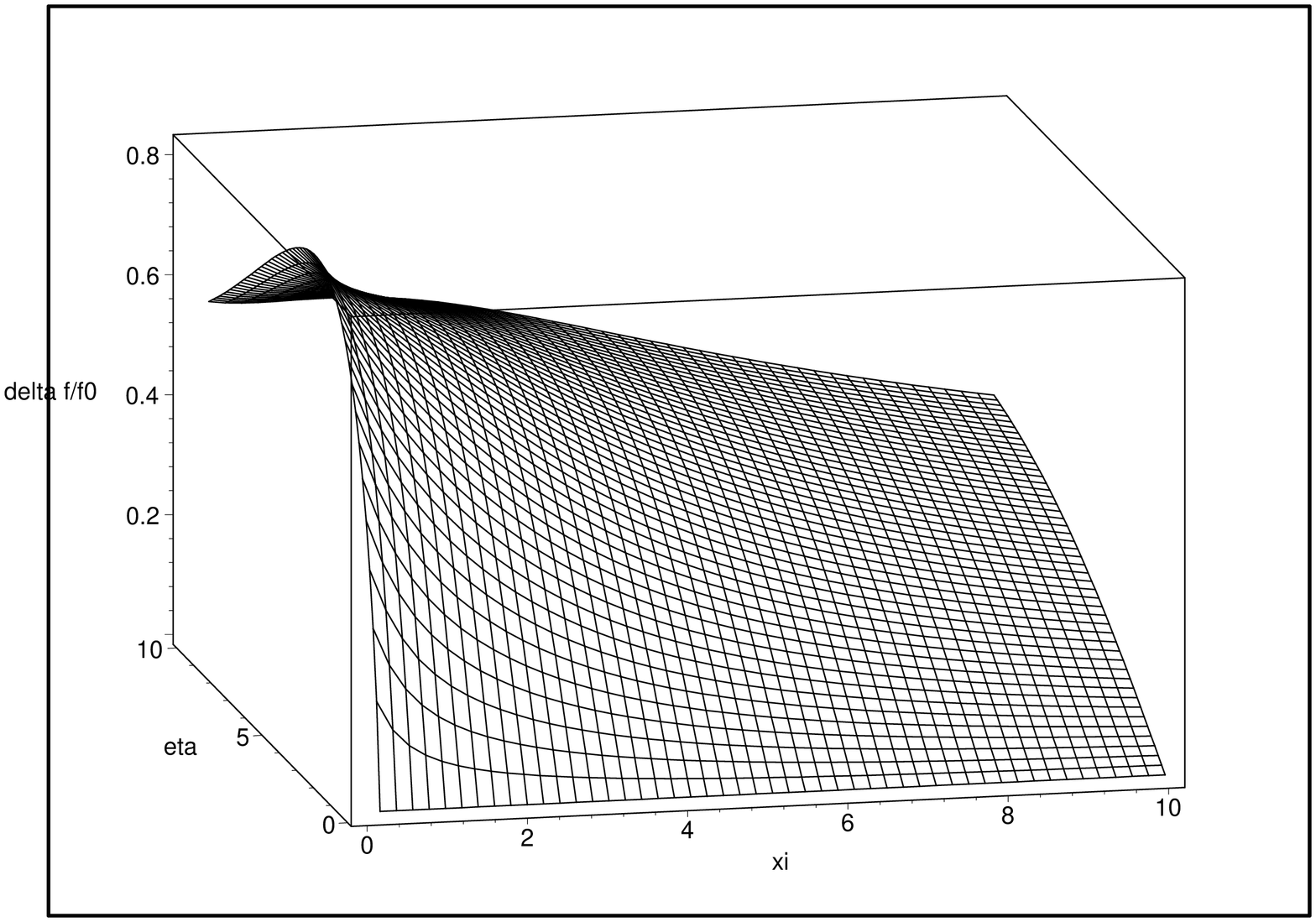,height=7cm,width=7cm,angle=-90}\label{delta-3}
\vskip 12pt \noindent {Fig.\bf \thefigure.}\hskip 12pt{\sl
Relative deviation of distribution function of X-bosons from
equilibrium $\delta f (\xi,\eta,\sigma)/f_0$ under
$\sigma=1$.\hfill}
\vskip 12pt\noindent%

From dduced results follows that:
\begin{enumerate}
\item Maximum of relative deviation of distribution function from equilibrium , ($\delta f/f_0$), falls at
lesser values of momentum coordinate $\xi \leq 1$;
\item On initial stages relative deviation of distribution
function from equilibrium grows with time $\eta$;
\item However at achievement certain critical sufficiently
great value of time relative deviation of distribution function
from equilibrium begins to decrease, - it associates with a fact,
that by this moment of time X-bosons start to decay;
\item At very great times weak maximum of relative deviation of distribution
function from equilibrium appears and starts to migrate to the
range of great values of momentum coordinate.
\end{enumerate}
\vskip 24pt\noindent \refstepcounter{figure}%\setcounter{figure}{1}
\epsfig{file=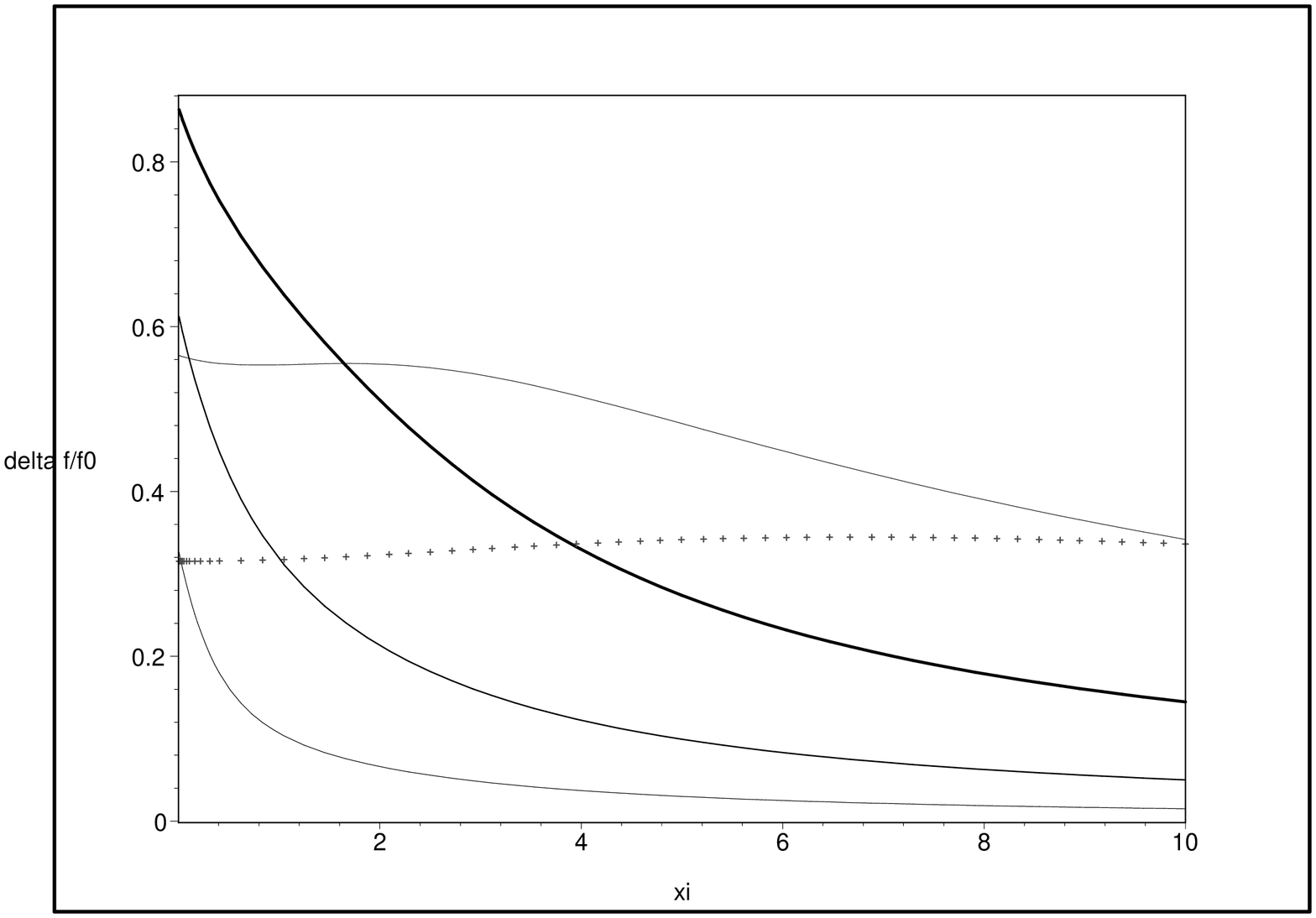,height=7cm,width=7cm,angle=-90}\label{delta-4}
\vskip 12pt \noindent {Fig.\bf \thefigure.}\hskip 12pt{\sl
Evolution of relative deviation of distribution function of
X-bosons from equilibrium\\
$\delta f (\xi,\eta,\sigma)/f_0$ under $\sigma=1$. Thin line -
$\eta=0.3$, normal line - $\eta=1$, heavy line - $\eta=3$, thin
dotted line - $\eta=10$, heavy dotted line - $\eta=20$. \hfill}
\vskip 12pt\noindent%

Let us notice that specific features of behavior of distribution
function's deviation were exhibit but were not explored in paper
\cite{Ignat4}. It is necessary to allow that under great values of
time coordinate $\eta$, absolute magnitude of density of X-bosons
number becomes vanishingly small. On Fig. \ref{delta-2} -
\ref{delta_0} are shown results of numeric integration of
expression (\ref{III.121_new}) for density of X-bosons deviation
from equilibrium, $\xi^2\delta f$, subject to time, $\eta$, and
momentum, $\xi$ for values $\sigma=0.3, 1, 3$.

\vskip 24pt\noindent \refstepcounter{figure}%\setcounter{figure}{1}
\epsfig{file=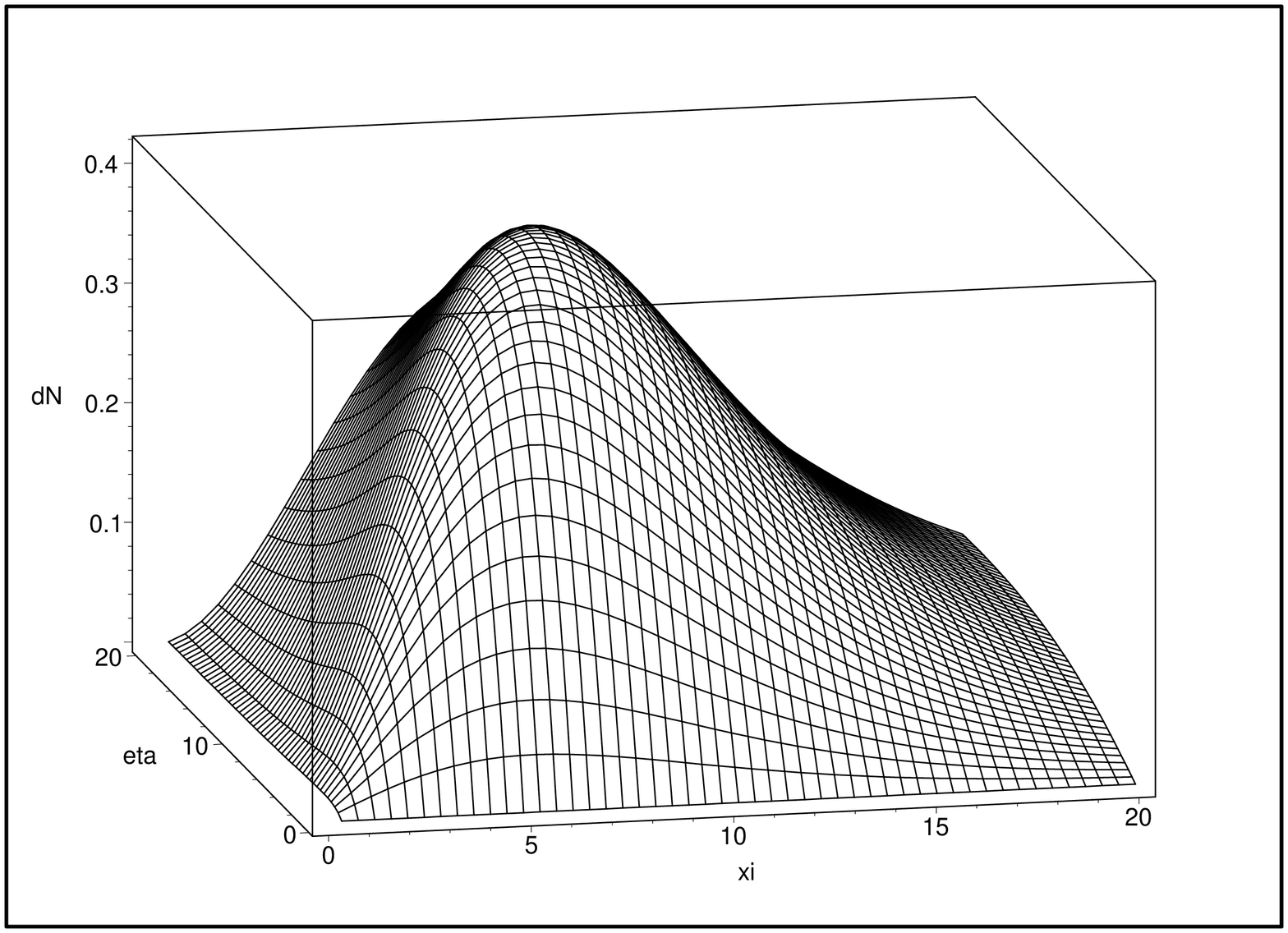,height=7cm,width=7cm,angle=-90}\label{delta-2}
\vskip 12pt \noindent {Fig.\bf \thefigure.}\hskip 12pt{\sl Density
of non-equilibrium X-bosons number $dN=\xi^2\delta f
(\xi,\eta,\sigma)$ under $\sigma=0,3$.\hfill}
\vskip 12pt\noindent%

\vskip 24pt\noindent \refstepcounter{figure}%\setcounter{figure}{1}
\epsfig{file=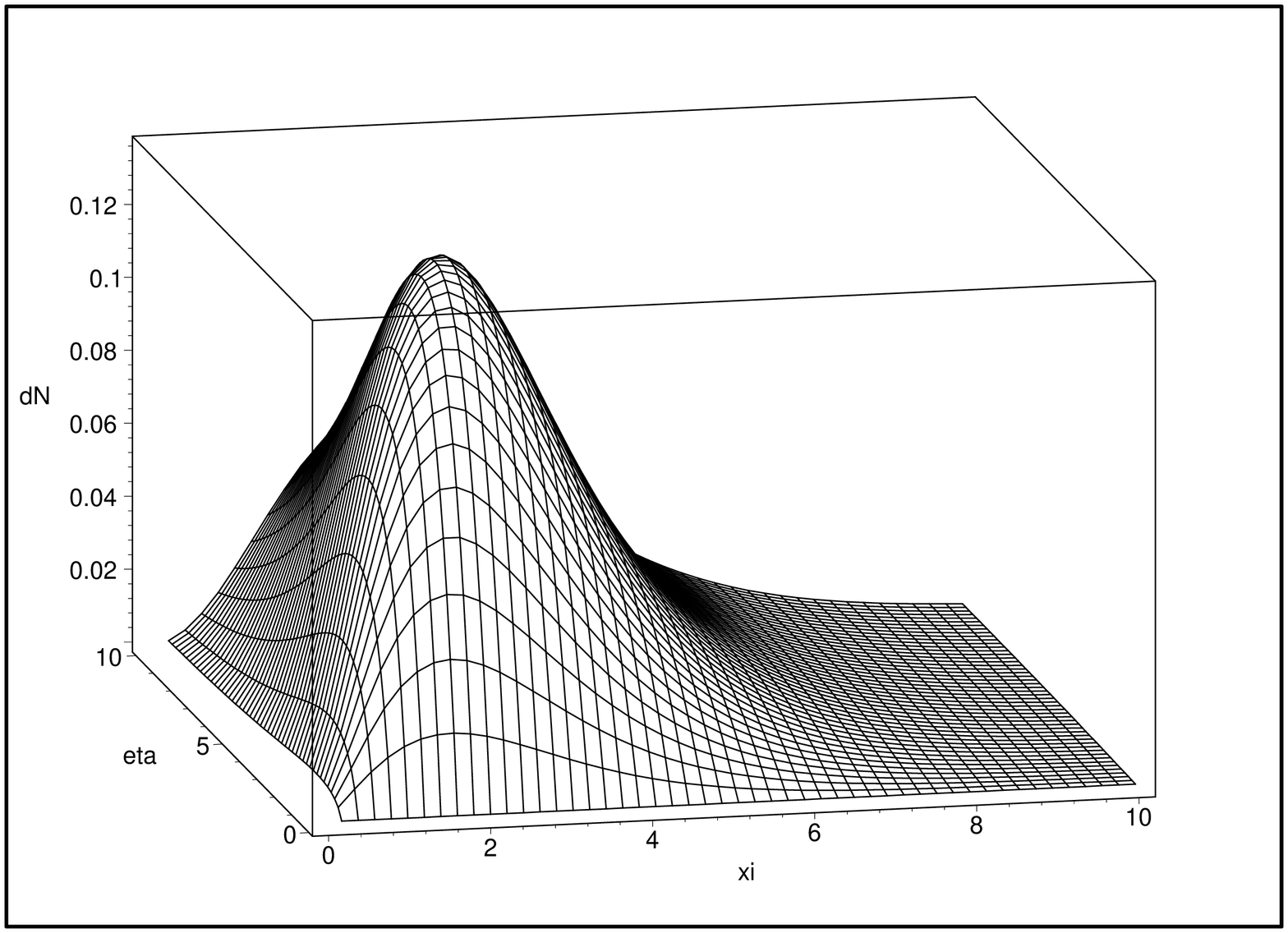,height=7cm,width=7cm,angle=-90}\label{delta-1}
\vskip 12pt \noindent {Fig.\bf \thefigure.}\hskip 12pt{\sl Density
of non-equilibrium X-bosons number $dN=\xi^2\delta f
(\xi,\eta,\sigma)$ under $\sigma=1$.\hfill}
\vskip 12pt\noindent%

\vskip 24pt\noindent \refstepcounter{figure}%\setcounter{figure}{1}
\epsfig{file=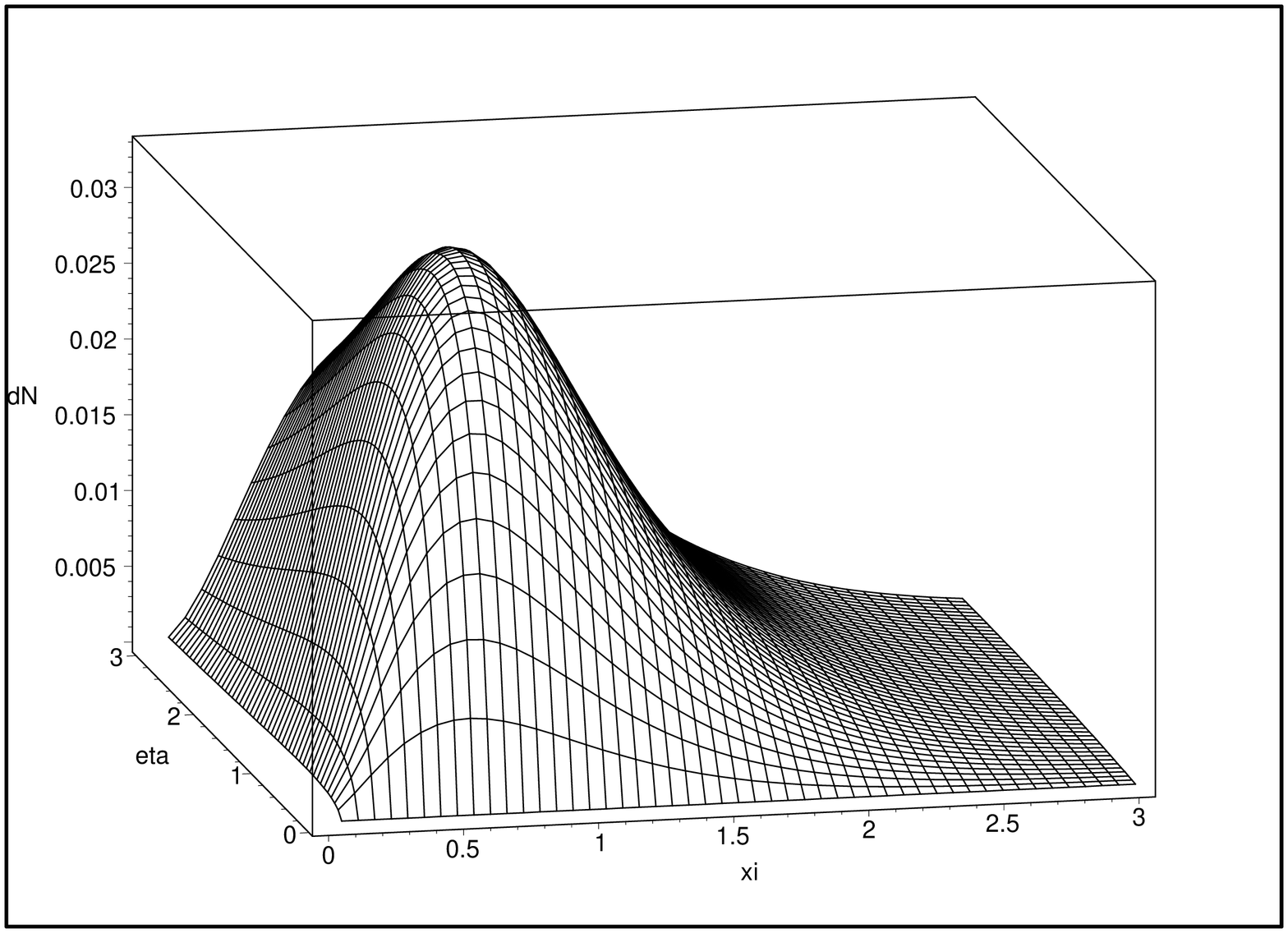,height=7cm,width=7cm,angle=-90}\label{delta_0}
\vskip 12pt \noindent {Fig.\bf \thefigure.}\hskip 12pt{\sl
ÏDensity of non-equilibrium X-bosons number $dN=\xi^2\delta f
(\xi,\eta,\sigma)$ under $\sigma=3$.\hfill}
\vskip 12pt\noindent%

On Fig. \ref{delta_1}- \ref{delta_3} are shown results of numeric
integration for evolution of X-bosons density's deviation from
equilibrium, $\xi^2\delta f$, calculated by formula
(\ref{III.121_new}), for values $\sigma=3, 1, 0.3$.

\vskip 24pt\noindent \refstepcounter{figure}%\setcounter{figure}{1}
\epsfig{file=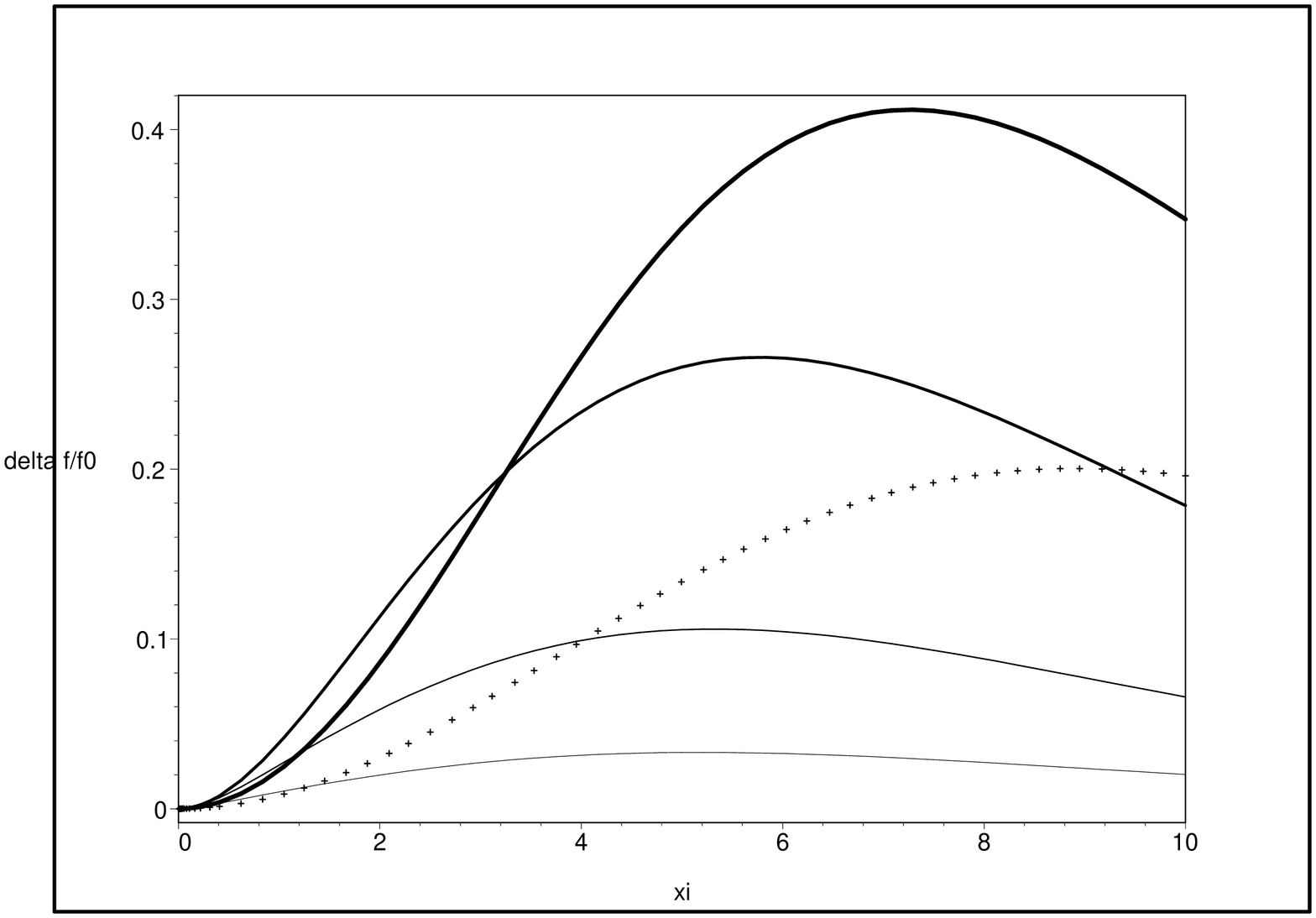,height=7cm,width=7cm,angle=-90}\label{delta_1}
\vskip 12pt \noindent {Fig.\bf \thefigure.}\hskip 12pt{\sl
Evolution of number density of non-equilibrium X-bosons
$dN=\xi^2\delta f (\xi,\eta,\sigma)$ under $\sigma=0.3$. Thin line
-$\eta=0.3,$ medium line - $\eta=1$, heavy line - $\eta=3$, the
heaviest line - $\eta=10$, dotted line - $\eta=30$.\hfill}
\vskip 12pt\noindent%
\vskip 24pt\noindent \refstepcounter{figure}%\setcounter{figure}{1}
\epsfig{file=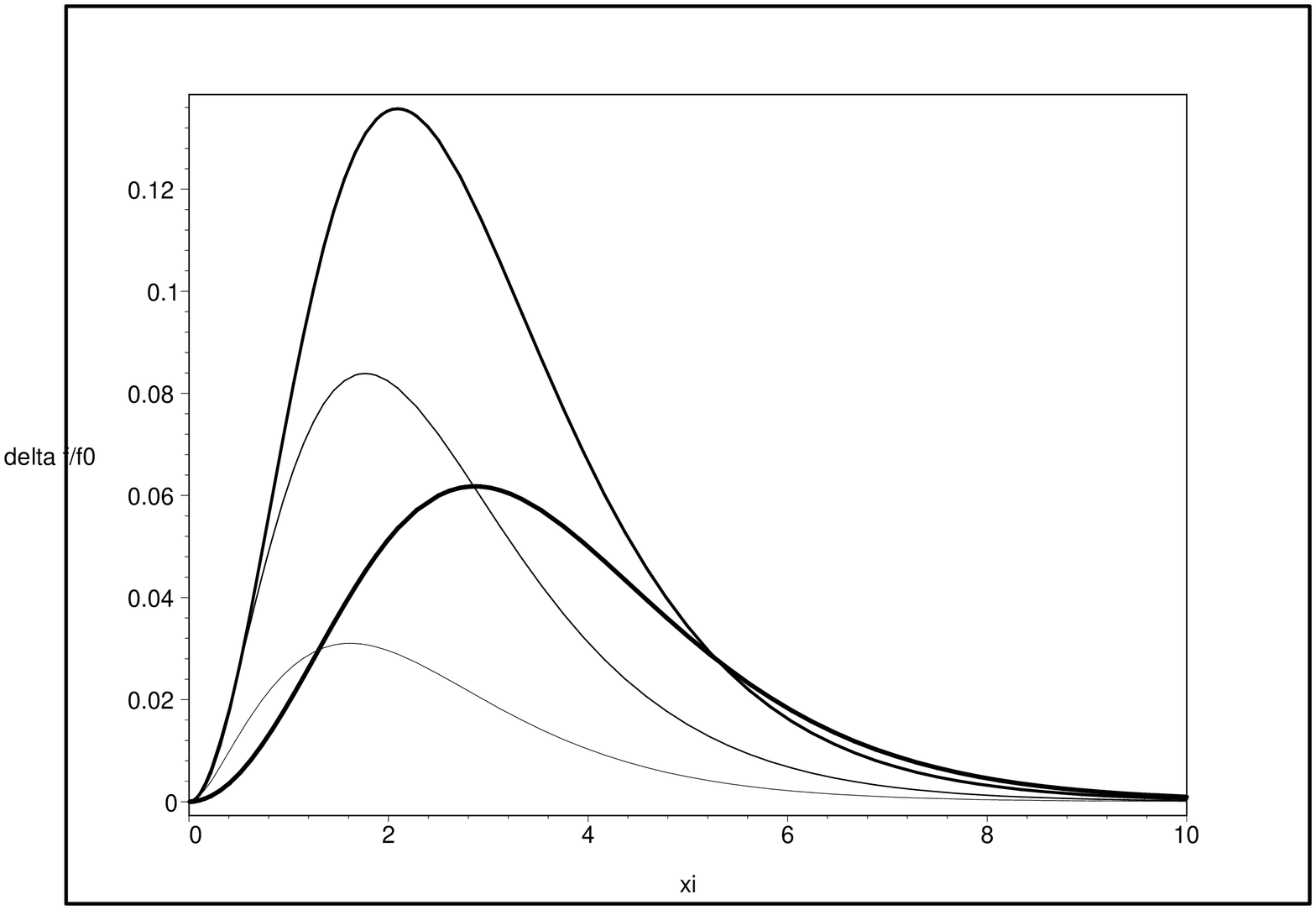,height=7cm,width=7cm,angle=-90}\label{delta_2}
\vskip 12pt \noindent {Fig.\bf \thefigure.}\hskip 12pt{\sl
Evolution of number density of non-equilibrium X-bosons
$dN=\xi^2\delta f (\xi,\eta,\sigma)$ under $\sigma=1$. Thin line
-$\eta=0.3,$ medium line ëèíèÿ - $\eta=1$, heavy line - $\eta=3$,
the heaviest line - $\eta=10$.\hfill}
\vskip 12pt\noindent%
\vskip 24pt\noindent \refstepcounter{figure}%\setcounter{figure}{1}
\epsfig{file=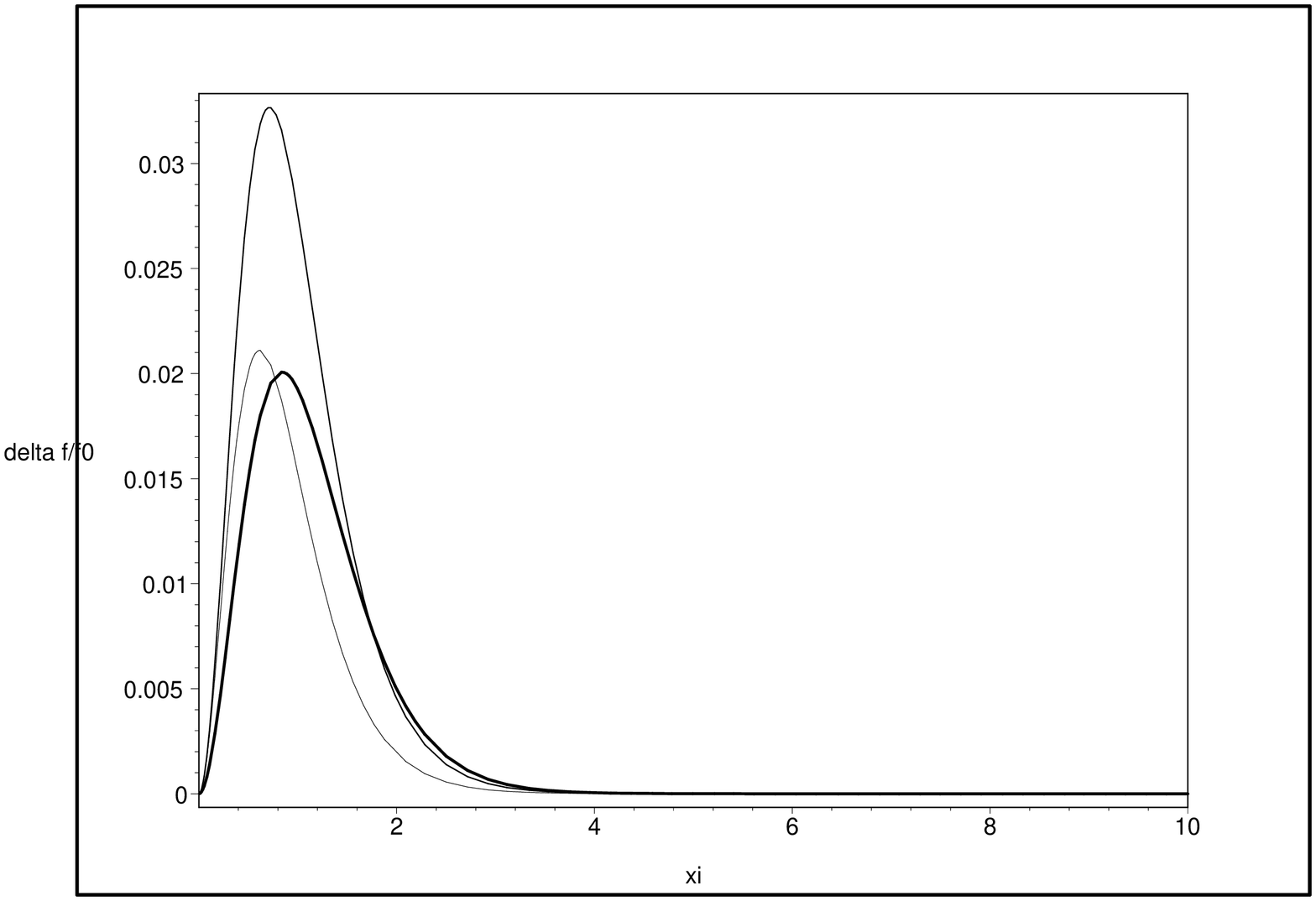,height=7cm,width=7cm,angle=-90}\label{delta_3}
\vskip 12pt \noindent {Fig.\bf \thefigure.}\hskip 12pt{\sl
Evolution of number density of non-equilibrium X-bosons
$dN=\xi^2\delta f (\xi,\eta,\sigma)$ under $\sigma=3$. Thin line
-$\eta=0.3,$ medium line - $\eta=1$, heavy line - $\eta=3$.\hfill}
\vskip 12pt\noindent%

\subsection{Exponential factor in formula (\ref{III.126})}
Further from formulas (\ref{III.126})-(\ref{III.128}) it is seen
that final solution defines not by function itself
$\Phi(\mathbb{P},t)$, but by it's time derivative, calculating
which we will find, proceeding to variables $\xi,\eta$ according
to formulas (\ref{yunew1}), (\ref{yunew2}):
\begin{equation}\label{yunew6}
\dot{\Phi}(\mathbb{P},t)=\frac{1}{\tau_0}
\frac{\sqrt{\eta}\beta_0(\eta,\sigma,\xi)}{\sqrt{\eta+\xi^2}}
\end{equation}
or, in Boltzmann approximation:
\begin{equation}\label{yunew6_1}
\dot{\Phi}(\mathbb{P},t)=\frac{1}{2\tau_0}
\frac{\sqrt{\eta}}{\sqrt{\eta+\xi^2}}
\end{equation}
Thus, calculating an integral by momentums we will receive
expression for function $\Psi(\eta,\sigma)$ (\ref{III.127}) in
Boltzmann approximation:
\begin{equation}\label{yunew7}
\Psi(\eta,\sigma)=\frac{N_X\sigma^2\eta}{\pi^2\tau_0}
K_1(\sigma\sqrt{\eta}),
\end{equation}
where $K_n(z)$ - modified Bessel function (Bessel imagunary
function) \cite{Lebed}:
\begin{equation}\label{K_n_1}
K_\nu(z)=\int\limits_0^\infty e^{-z\cosh u}\cosh \nu u du,
\end{equation}
at that following recurrence equations are correct:
\begin{equation}\label{K_n_2}
K_{\nu+1}(z)-K_{\nu-1}(z)=\frac{2\nu}{z}K_\nu(z).
\end{equation}

Thus, exponential factor in formula (\ref{III.126}) is equal:
\begin{equation}\label{yunew8}\Psi_e(\eta,\sigma)=\exp(-\int\limits_t^\infty
\Psi(t',\sigma)dt')=\exp\left(-\frac{2N_X}{\pi^2\sigma^2}\int\limits_{\sigma\sqrt{\eta}}^\infty
x^3K_1(x)dx\right).
\end{equation}

Considering that:
\begin{equation}\label{int_K1}
\int\limits_0^\infty x^3K_1(x)dx=\frac{3\pi}{2},
\end{equation}
it is not hard to receive limitary relation:
\begin{equation}\label{lim_psi}
\Psi_e(\eta,\sigma)\approx
\exp\left(-\frac{3N_X}{\pi\sigma^2}\right),\qquad
(\sigma\sqrt{\eta} \to 0),
\end{equation}
as well as asymptotic expansion of function $\Psi_e(\eta,\sigma)$
at small values of argument $\sigma\sqrt{\eta}$:
\begin{equation}\label{asimp_psi}
\Psi_e(\eta,\sigma)\approx
\exp\left(-\frac{3N_X}{\pi\sigma^2}+\frac{2N_X\sigma \eta^{3/2}
}{3\pi^2}\right),\quad (\sigma\sqrt{\eta} \to 0).
\end{equation}

In other limitary case accounting asymptotic expansion of Bessel
functions at great values of argument we will receive:

$$\Psi_e(\eta,\sigma)\approx
\exp\left(-\frac{3N_X}{\pi^2}\sqrt{2\pi\sigma}\eta^{5/4}\rm{e}^{-\sigma\sqrt{\eta}}\right)
$$
\begin{equation}\label{lim_psi1}
\approx
1-\frac{3N_X}{\pi^2}\sqrt{2\pi\sigma}\eta^{5/4}\rm{e}^{-\sigma\sqrt{\eta}}
\qquad (\sigma\sqrt{\eta}\to\infty) .
\end{equation}

On Fig.\ref{int_psi} dependence of exponential factor from
variables $\eta, \sigma$, received by numerical integration for
$N_X=1$, is shown.

\vskip 24pt\noindent \refstepcounter{figure}%\setcounter{figure}{1}
\epsfig{file=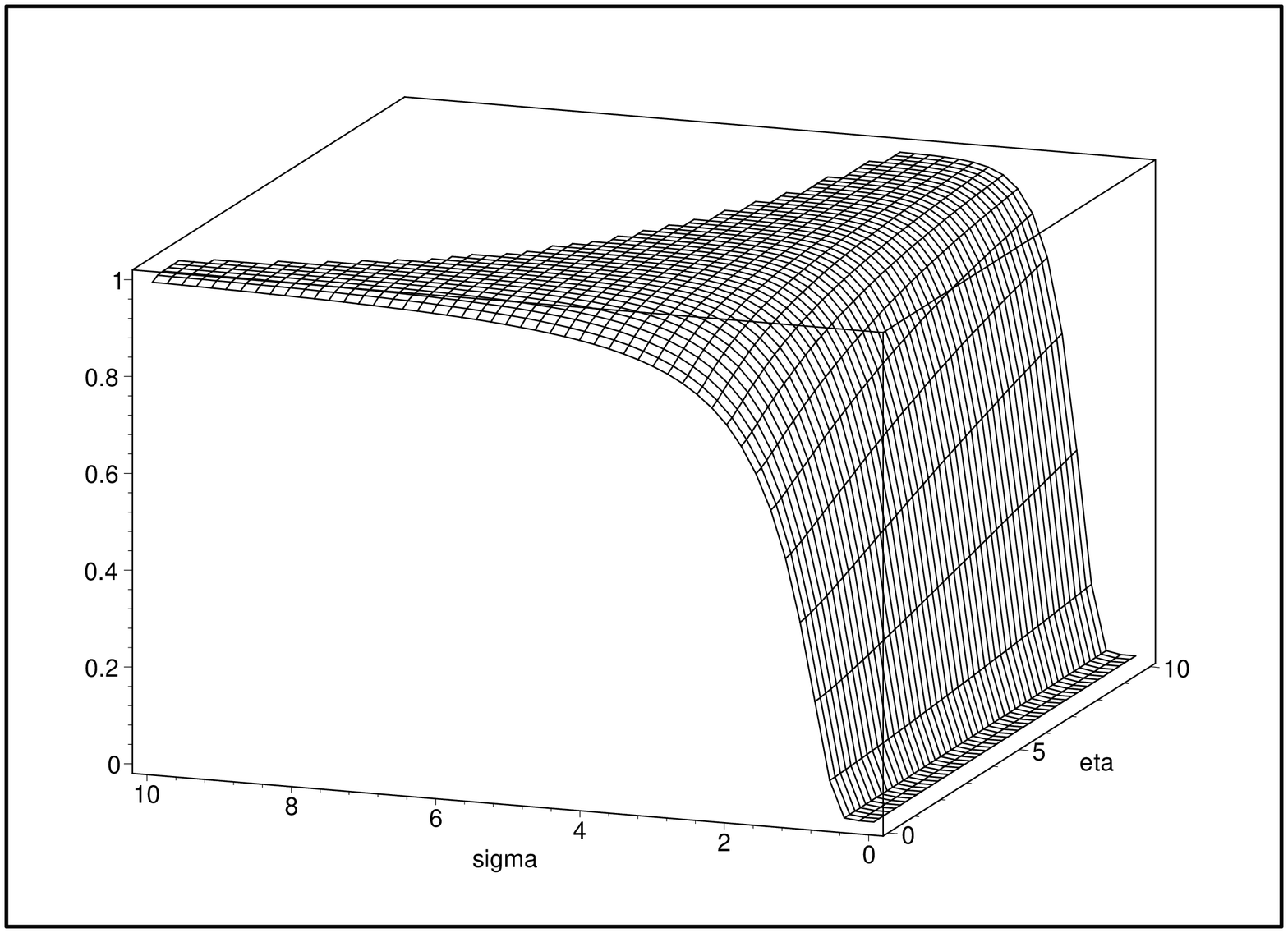,height=7cm,width=7cm,angle=-90}\label{int_psi}
\vskip 12pt \noindent {Fig.\bf \thefigure.}\hskip 12pt{\sl
Function $\Psi_e(\eta,\sigma)=\exp(-\int\limits_t^\infty\Psi dt'$)
at $N_X=1$. \hfill}
\vskip 12pt\noindent%

On Fig. \ref{psi1t} is shown an evolution of exponential factor
$\Psi_1(\eta,\sigma)$ at various values of parameter $\sigma$,

\vskip 24pt\noindent \refstepcounter{figure}%\setcounter{figure}{1}
\epsfig{file=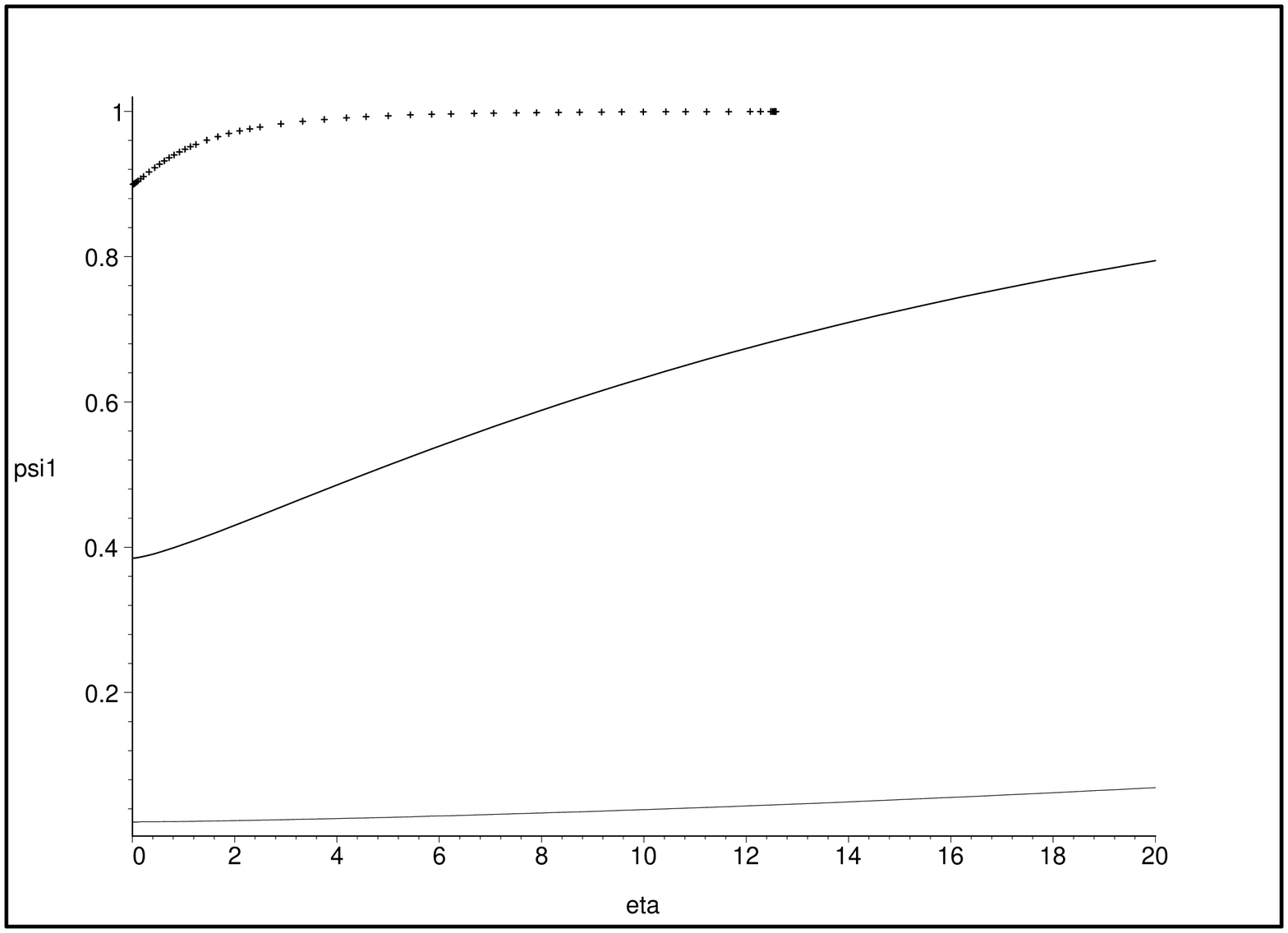,height=7cm,width=7cm,angle=-90}\label{psi1t}
\vskip 12pt \noindent {Fig.\bf \thefigure.}\hskip 12pt{\sl
Function $\Psi_e(\eta,\sigma)=\exp(-\int\limits_t^\infty\Psi dt'$)
at $N_X=1$. Thin line: $\sigma=0.3$, medium line: $\sigma=1$,
dotted line: $\sigma=3$. \hfill}
\vskip 12pt\noindent%
On Fig. \ref{psi1t_s} influence of number of X-boson types on
exponential factor, $N_X$ is shown.
\vskip 24pt\noindent \refstepcounter{figure}%\setcounter{figure}{1}
\epsfig{file=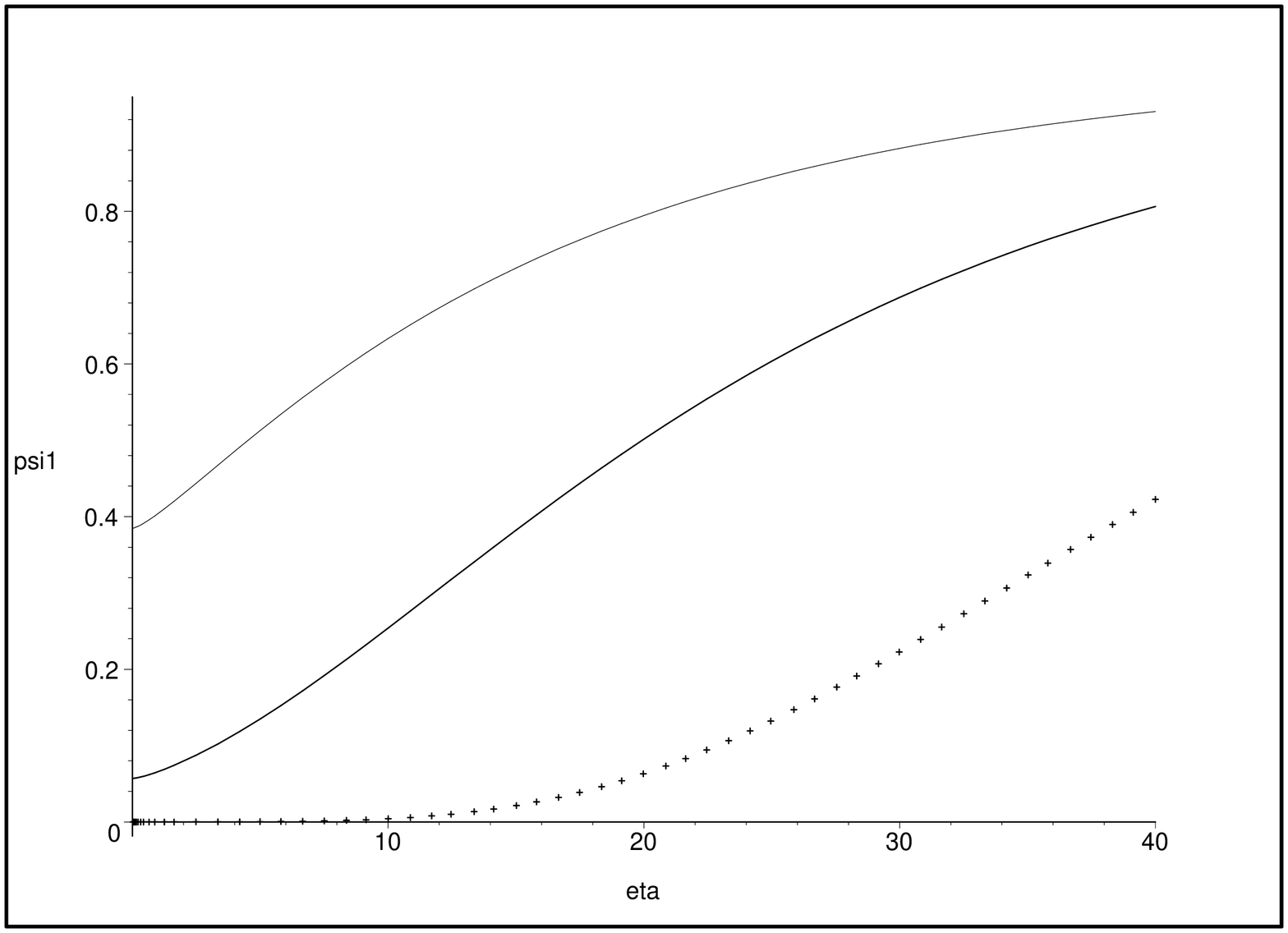,height=7cm,width=7cm,angle=-90}\label{psi1t_s}
\vskip 12pt \noindent {Fig.\bf \thefigure.}\hskip 12pt{\sl
Influence of number of X-boson types, $N_X$, on exponential factor
$\Psi_1(\eta,\sigma)=\exp(-\int\limits_t^\infty\Psi dt'$) at
$\sigma=1$. Thin line: $N_X=1$, heavy line: $N_X=3$, dotted line:
$N_X=12$. \hfill}
\vskip 12pt\noindent%

On Fig. \ref{psi1t_s} are shown values of exponential factor,
calculated by numerical methods and using asymptotic estimates
(\ref{asimp_psi}), (\ref{lim_psi1}).

\vskip 24pt\noindent \refstepcounter{figure}%\setcounter{figure}{1}
\epsfig{file=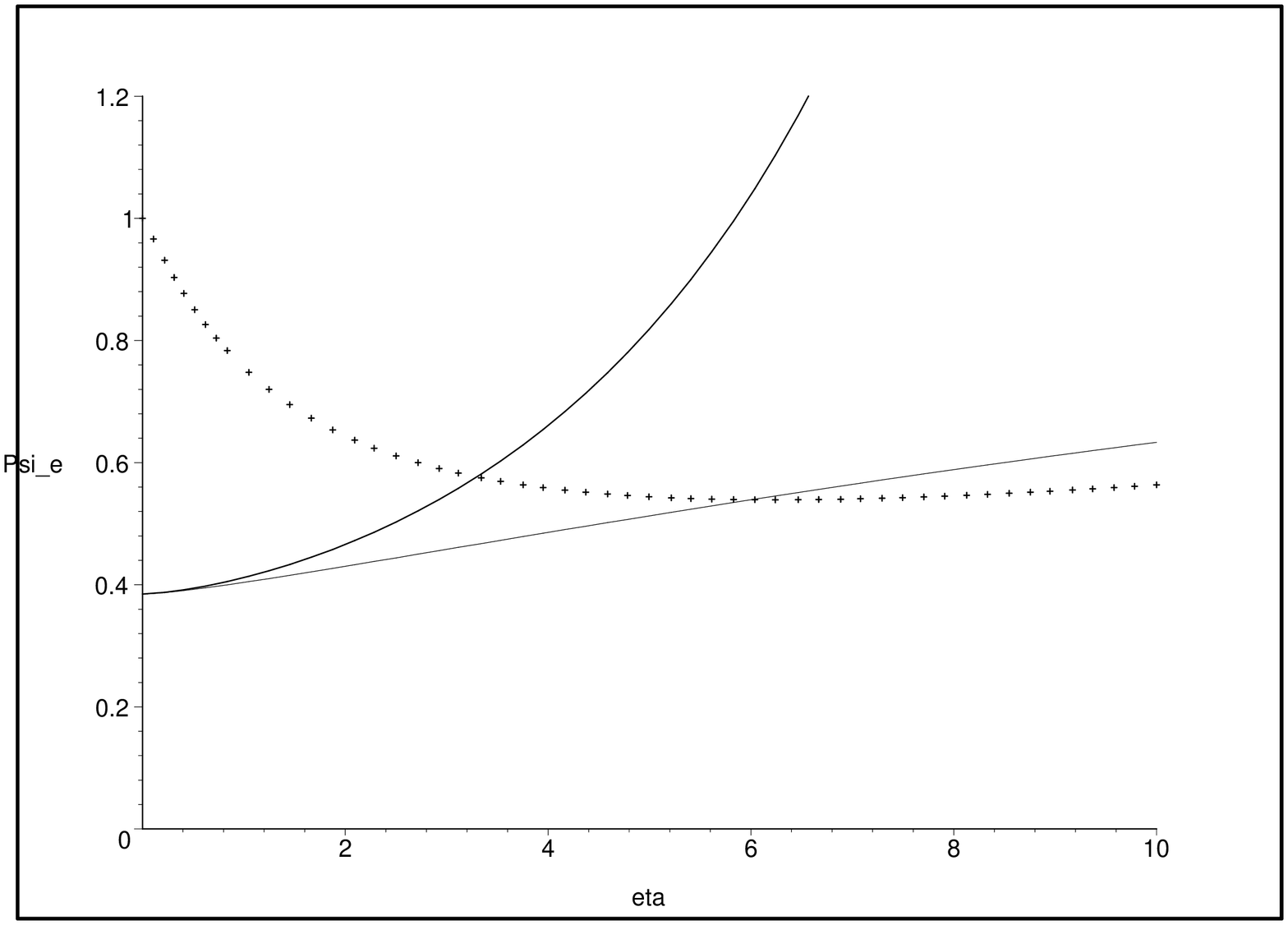,height=7cm,width=7cm,angle=-90}\label{psi1t_s}
\vskip 12pt \noindent {Fig.\bf \thefigure.}\hskip 12pt{\sl
Comparison of exponential factor values, calculated by numerical
methods (thin line) and by asymptotic estimates (\ref{asimp_psi})
-heavy line and (\ref{lim_psi1}) - dotted line. \hfill}
\vskip 12pt\noindent%

\end{document}